\documentclass[epjCONF]{svjour}
\usepackage{graphics}
\usepackage[varg]{txfonts} % Times fonts
\usepackage[latin1]{inputenc}
\usepackage{color}

\session-title{Hot and Cold Baryonic Matter -- HCBM 2010}
\begin{document}
\title{QCD Green's Functions and Phases of  Strongly-Interacting Matter }
\author{Reinhard
Alkofer\fnmsep\thanks{\email{reinhard.alkofer@uni-graz.at}} 
\and 
Mario Mitter\fnmsep\thanks{\email{mario.mitter@uni-graz.at}} 
\and 
Bernd-Jochen Schaefer\fnmsep\thanks{\email{bernd-jochen.schaefer@uni-graz.at}}}
\institute{Institut f\"ur Physik, Karl-Franzens--Universit\"at Graz,  
Universit\"atsplatz 5, 8010 Graz,
Austria}
\abstract{ After presenting a brief summary of functional approaches
  to QCD at vanishing temperatures and densities the application of
  QCD Green's functions at non-vanishing temperature and vanishing
  density is discussed. It is pointed out in which way the infrared
  behavior of the gluon propagator reflects the (de-)\-confinement
  transition. Numerical results for the quark propagator are given
  thereby verifying the relation between (de-)\-confinement and
  dynamical chiral symmetry breaking (restoration). Last but not least
  some results of Dyson-Schwinger equations for the
  color-superconducting phase at large densities are shown. }
\maketitle
\section{Introduction: Why Green's functions?}
\label{intro}

The QCD Green's functions, and thereby especially the ones in the
Landau gauge, have been in the focus of many recent investigations. In
principle, they provide a complete description of the Strong
Interaction. This implies that they also embody the non-perturbative
phenomena of QCD, a\-mongst them most prominently confinement,
dynamical chiral symmetry breaking, and the axial anomaly. On the
other hand, they serve as input into hadron phenomenology based on
bound-state equations: mesons and their properties are studied from
solutions of Bethe-Salpeter, and baryons from the solutions of
covariant Faddeev equations.\footnote{A recent example of such
  calculations can be found in
  \cite{Nicmorus:2010sd,SanchisAlepuz:2010in} and references therein.}

Of course, this raises immediately the question whether Green's
functions are suitable to study the properties of the different QCD
phases and the phase diagram of QCD. Although this question will
likeley be answered affirmatively, the investigations, as described in
the following, are definitely at an early stage and require several
improvements before they can be considered conclusive. Nevertheless,
they are certainly not only very promising but also the best access to
map out the QCD phase diagram.

Before continuing with a brief summary of functional approaches to QCD  at
vanishing temperatures and densities, we want to point out that studies of the
QCD phase diagram within functional methods have been discussed from a
broader perspective than here in \cite{Pawlowski:2010ht} (see also references 
therein).

\section{Functional Approaches to QCD}
\label{secFun}

As stated above, QCD Green's functions are a promising powerful tool.
However, in order to determine them in continuum quantum field theory,
in principle, an infinite hierarchy of coupled complicated equations
has to be solved. As for QCD, we are of course most interested in
their infrared behaviour, {\it i.e.}, in the Green's functions in the
stron\-gly-interacting domain. To this end it turns out that
restricting oneself to the primitively divergent Green's functions,
which are the most interesting ones, provides an appropriate starting
point (see {\it e.g. \/} \cite{Alkofer:2008bs} and references
therein). Even in the Landau gauge with the least number of
primitively divergent functions a formidable task is left: there are
seven primitively divergent functions while in Landau gauge Yang-Mills
theory there are still five primitively divergent functions, namely
the gluon and ghost propagators as well as the 3-gluon, 4-gluon and
the gluon-ghost vertices. Including quarks, one has to consider in
addition the quark propagator and the quark-gluon vertex.

The functional approaches include methods which are based on the
Dyson-Schwinger Equations (DSEs), the Functional Renormalisation Group
Equations (FRG), and the $n$-Particle Irredu\-cible Actions (nPI).
Hereby, the DSEs, being the equations of motion for the Green's
functions, are derived straightforwardly from the generating
functional. The FRG is based on the idea of employing
energy-momen\-tum cutoffs to ``integrate out'' the high-momentum modes
and the nPI are effective actions allow to derive symmetry-preserving
equations for the Green's functions.

The advantages of functional methods are: \\[-5mm]
\begin{itemize}
\item It is straightforward to implement chiral fermions and Goldstone's
theorem.
\item There exist analytical infrared solutions.
\item Hadrons are described in terms of their fundamental substructure.
\item There is no sign problem at non-vanishing chemical potential. 
\end{itemize}
Compared to the most widely employed non-perturbative first-principle
method, the lattice gauge
theory, these advantages are clear benefits. But these do not come for free,
 because, on the other hand, 
functional methods miss the main advantages of lattice gauge theory:\\[-5mm]
\goodbreak
\begin{itemize}
\item There are no truncations.
\item Manifest gauge invariance.
\end{itemize}
This makes plain that, whenever possible, an intelligent combinations of methods
will provide the most reliable results.

\section{What do we know for $T=0$ and $\mu =0$?}

\subsection{Infrared Structure of Landau gauge Yang-Mills theory}

\begin{figure}
\centerline{
\resizebox{0.9\columnwidth}{!}{\includegraphics{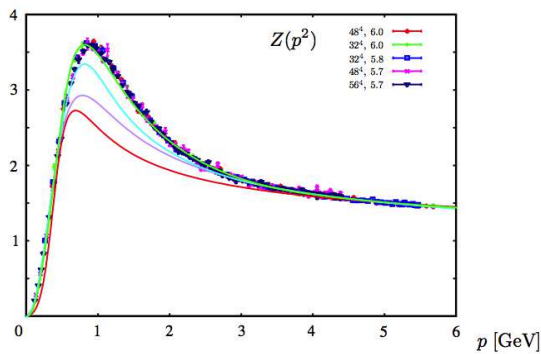}}
}
\caption{The Landau gauge gluon renormalization function obtained from
  different calculations, see text for details.}
\label{GluonProp}       
\end{figure}

In order to demonstrate the progress of functional methods over the
last decade it is interesting to compare different steps in the
calculation of the Landau gauge gluon propagator with corresponding
lattice results. Many lattice calculations of the gluon propagator are
available, and it would be beyond the scope of these proceedings to
cite only the major steps in improving on these calculations. In Fig.\
\ref{GluonProp} we have chosen the lattice results for the gluon
renormalization function (see also Eq.~(\ref{ScalProp})) of Ref.\
\cite{Sternbeck:2006cg} for comparison, but any other recent
calculation would equally well be suitable. The lowest lying line
displays the results of Ref.\ \cite{von Smekal:1997is} in which the
coupled DSEs for the gluon and ghost propagators have been solved
employing some approximative treatment of the involved angular
integrals. The second lowest curve is from Ref.\
\cite{Fischer:2002hna} in which the angular integrals could be treated
without any approximation. The curve above is the result of Ref.\
\cite{Pawlowski:2003hq} where the FRG has been used to solve for the
propagators. The highest lying curve which becomes also closest to the
lattice data is a result obtained by Jan Pawlowski based on the work
of Ref.\ \cite{Fischer:2008uz} where DSE and FRG methods have been
combined.\footnote{We thank Jan Pawlowski for discussing with us his
  work in preparation and Lorenz von Smekal for a compilation of the
  data (see also his talk under
  {http://www.thphys.uni-heidel\-berg.de/$\sim$smp/view/Main/Delta10Programme}).}

How has this success of functional methods been a\-chieved? To this
end one should note that the ghost-gluon vertex becomes bare in the
limit of vanishing ghost momentum due to the trans\-versality of the
gluon propagator. This has been noticed already 40 years ago
\cite{Taylor:1971ff} and has been tested in several non-perturbative
investigations
\cite{Lerche:2002ep,Cucchieri:2004sq,Schleifenbaum:2004id}. With the
knowledge that the ghost-gluon vertex stays regular in the infrared
one can now analyze the ghost propagator DSE. Power-law like ans\"atze
for the gluon and ghost renormalization functions lead to a relation
between the infrared exponents \cite{von Smekal:1997is} (this kind of
relations are nowadays known as scaling relations). Denoting by
$Z(p^2)$ and $G(p^2)$ the gluon and the ghost renormalization
function, respectively, one obtains for $p^2\to 0$
\begin{equation}
Z(p^2) \sim (p^2)^{2\kappa}, \qquad G(p^2) \sim (p^2)^{-\kappa}, 
\label{ScalProp}
\end{equation}
in terms of the ghost propagator
infrared exponent $\kappa$. As it can be shown that 
 $0.5 <\kappa < 1$
\cite{Watson:2001yv} the gluon propagator is infrared suppressed
and the ghost propagator is infrared enhanced.\footnote{It is an interesting
side remark that with this behaviour the Kugo--Ojima confinement criterion, the 
Oehme--Zimmermann superconvergence relation, and the Gribov--Zwanziger horizon
condition are fulfilled, see, {\it e.g.}, the review \cite{Alkofer:2000wg} for
a detailed discussion.}

This type of infrared analysis of DSEs can be extended to all
Yang-Mills vertex functions
\cite{Alkofer:2004it,Huber:2007kc}.\footnote{Note that a recently
  published MATHEMATICA code can be used to simplify significantly the
  derivation of the corresponding DSEs \cite{Alkofer:2008nt}.}
Employing in addition the FRG equations, and requiring that these two,
seemingly different, towers of equations have to give identical
Green's functions, allows one to restrict the type of solutions
strongly. There is one unique scaling solution with power laws for the
Green's functions \cite{Fischer:2006vf,Fischer:2009tn} and a
one-parameter family of solutions, the so-called decoupling solutions.
At this point some notes are in order. Quite some time ago, in the
Coulomb gauge, a similar situation has been found in variational
approaches \cite{Szczepaniak:2001rg}: A family of infrared trivial
solution possesses as an endpoint a critical solution characterized by
some infrared power laws. As it concerns the Landau gauge: Numerical
solutions of the decoupling type (there called ``massive solution'')
have been published in \cite{Aguilar:2008xm,Boucaud:2008ky} and
references therein. A recent detailed description and comparison of
these two type of solutions has been given in Ref.\
\cite{Fischer:2008uz}.\footnote{The infrared analysis for both type of
  solutions is described in Refs. \cite{Alkofer:2008jy,Huber:2009wh}.}
Firstly, one has to note that the scaling solution respects the BRST
symmetry whereas the decoupling solutions break this symmetry
\cite{Fischer:2008uz}. Secondly, almost all lattice calculations of
the gluon propagator favor a decoupling solution. But lattice studies
at strong coupling
\cite{Sternbeck:2008mv,Maas:2009ph,Cucchieri:2009zt} reveal the
existence of a regime where the scaling relation between the gluon and
the ghost propagator is fulfilled, and the corresponding infrared
exponent $\kappa$ is very close to the value determined in truncated
continuum studies with $\kappa=0.595$. A potential resolution of this
puzzle has been offered recently in Ref. \cite{Maas:2009se}: The
infrared behaviour of the Green's function depends on the
non-perturbative completion of the gauge.

At this point it has to be emphasized that this discussion is of fundamental
importance (especially with respect to the role of the BRST symmetry in a
non-perturbative approach) but the difference between the two types of solutions
is phenomenologically irrelevant.\footnote{Maybe with the exception of how to
describe the axial anomaly with Green's functions, see below.}
 
For completeness we will give the infrared behaviour of all one-particle
irreducible Green's functions in the scaling solution in the simplified case with
only one external scale $p^2\to 0$. For a function with $n$ external
ghost and antighost as well as $m$ gluon legs one has:
\begin{equation}
\Gamma^{n,m}(p^2) \sim (p^2)^{({n-m})\kappa} .
\end{equation}
Note that this solution fulfills all DSEs, all FRG equations and all
Slavnov-Taylor identities. In addition, it verifies the hypothesis of infrared
ghost dominance \cite{Zwanziger:2003cf} and leads to infrared diverging 3- and
4-gluon vertex functions. 

As a further side remark we want to add that the gluon propagator
violates positivity \cite{Alkofer:2003jj,Bowman:2007du} which is a property
related to the confinement of transverse gluons. Furthermore, in Ref.
\cite{Alkofer:2003jj} an analytic structure for the gluon propagator has been
proposed in which the gluon propagator is analytic in the complex $p^2$-plane
except the space-like real half-axis.

\subsection{Quarks: Confinement vs.~D$\chi$SB and the U$_A$(1) anomaly }

Independent of the type of the solutions -- scaling or decoupling --
the gluon propagator is infrared suppressed, and therefore quark
confinement cannot be generated by any type of gluon exchange together
with an infrared-bounded quark-gluon vertex. To add quarks one
considers the DSE for the quark propagator together with the one for
the quark-gluon vertex in a self-consistent way
\cite{Alkofer:2006gz,Alkofer:2008tt}. Hereby, a significant difference
of the quarks compared to the Yang-Mills fields has to be taken into
account: Since they possess a mass, and since dynamical chiral
symmetry breaking (D$\chi$SB) does occur, the quark propagator will
always approach a constant in the infrared.

In general, the fully dressed quark-gluon vertex can be expanded in twelve linearly
independent Dirac tensors. Half of the coefficient functions would vanish if
chiral symmetry were realized in the Wigner-Weyl mode. From a solution of the
Dyson-Schwinger equations we infer that  these {\em ``scalar''} structures are,
in the chiral limit, generated non-perturba\-tively together with the dynamical
quark mass function in a self-consistent fashion. Thus, dynamical chiral symmetry
breaking manifests itself not only in the propagator but also in the quark-gluon
vertex.

Based on the Yang-Mills scaling solution one obtains from an infrared analysis
an infrared divergent solution for the quark-gluon vertex such that the Dirac
vector and {``scalar''} components of this  vertex are infrared divergent with
an exponent $-\kappa - \frac 1 2$ if either all momenta \cite{Alkofer:2006gz} or
the gluon momentum vanish \cite{Alkofer:2008tt}. A numerical
solution of a truncated set of DSEs confirms this infrared behaviour. The
essential components, to obtain this solution, are the scalar Dirac amplitudes of
the quark-gluon vertex and the scalar part of the quark propagator. Both are
only present when chiral symmetry is broken, either explicitely or dynamically.

This self-consistent quark propagator and quark-gluon vertex solution
relates to quark confinement via the anomalous infrared exponent of
the four-quark function. The static quark potential can be calculated
from this four-quark one-particle irreducible Green function, which
behaves like\\ $(p^2)^{-2}$ for $p^2\to0$ due to the infrared
enhancement of the quark-gluon vertex for vanishing gluon momentum.
Using the well-known relation for a function $F\propto (p^2)^{-2}$ one
gets
\begin{equation}
V({\bf r}) = \int \frac{d^3p}{(2\pi)^3}  F(p^0=0,{\bf p})  e^{i {\bf p r}}
\ \ \sim \ \ |{\bf r} |
\end{equation}
for the static quark-antiquark potential $V({\bf r})$. Therefore the
infrared divergence of the quark-gluon vertex, as found in the scaling
solution of the coupled system of DSEs, the vertex overcompensates the
infared suppression of the gluon propagator such that one obtains a
linearly rising potential.

Generally, if an approach is complete it has to describe all phenomena of a
given theory. This especially implies that also the axial U$_A$(1) anomaly
should be described with the Green's functions, or in other words, that the QCD
Green's functions incorporate the topological susceptibility of the QCD vacuum.
To this end we would like to point out that the infrared divergence of the
quark-gluon vertex, mentioned above, indeed generates an $\eta^\prime$ mass and
therefore describes the axial anomaly \cite{Alkofer:2008et}. Following an old
idea \cite{Kogut:1973ab} which attributed the $\eta^\prime$ mass to the infrared
slavery of quarks it was shown that the infrared divergence of the quark
four-point function in the soft limit is exactly the one which is needed to find
a non-vanishing $\eta^\prime$ mass. However, the appearence of the correct
infrared singularity in single Feynman diagrams is the case only for the
scaling solution. To our best knowledge, it is up to now unknown how the axial
anomaly is encoded in the elementary Green's functions of the decoupling
solution as only a resummation of infinitely many diagrams will be able to
describe the anomaly when employing these solutions for the Green's functions.

\section{Phases of strongly-interacting matter:
How to go to $T\not = 0$ and $\mu \not =0$ ?
}

The exploration of the QCD phase diagram, in particular the higher
baryon-density regime, within functional approaches is a very active
field of research and of great importance for a deeper understanding
of the experimental data of the running and planned heavy-ion
programs. As already argued, in the last years a fruitful interaction
between the different functional methods has lead to a largely
quantitative understanding of QCD at vanishing temperature and density
while an understanding of the confinement mechanism and its relation
to the D$\chi$SB is not yet settled. At finite temperature and
chemical potential the situation is even much less clear. On the one
hand, with the help of functional techniques results for the pure
Yang-Mills sector of QCD at finite temperature as well as for the
hadronic sector of QCD could be obtained but on the other hand a full
QCD investigation is hampered by the fact that the gauge sector, {\it
  i.e.\/} the (de-)confinement transition, is not fully resolved.
Furthermore, for full QCD with dynamical quarks the (de-)confinement
transition is expected to be a smooth crossover as the underlying
center symmetry of the SU(N$_c$) gauge group is explicitly broken by
the quarks. Similarly, the nature of the chiral phase transition
depends, among other quantities, on the values of the current quark
masses which again break chiral symmetry explicitly. It is an
important observation that for small chemical potentials both phase
transitions lie with respect to the temperature remarkably close to
each other. This is interesting since the (de-)confinement transition
is driven by the gluodynamics while the chiral transition is govern by
strongly-interacting quarks. The deeper understanding of this
interrelation is a primary focus of the application of functional
approaches, see also \cite{Pawlowski:2010ht,Braun:2009gm}.

A first step towards full QCD with functional methods at non-vanishing
temperature and densities is done in the framework of effective models
which are generally constructed from non-perturbative Yang-Mills
effective potentials and hadronic potentials. Pure Yang-Mills theory
corresponds to the heavy-quark limit of QCD where the expectation
value of the Polyakov loop serves as an order parameter for the
(de-)confinement transition. In these models some information about
the confining glue sector of QCD is incorporated in an effective
Polyakov-loop potential that is extracted from pure Yang-Mills lattice
simulations at vanishing chemical potential. This sector is
effectively combined with the chiral quark(-meson) matter sector
leading to the Polyakov-NJL-type effective chiral models such as the
PNJL or PQM models, {\it e.g.\/},
\cite{Ratti:2005jh,Schaefer:2007pw,Schaefer:2009ui}. The matter sector
of these models has been studied also beyond mean-field level by
taking into account the quark-meson quantum fluctuations within a FRG
approach, for recent studies see Refs
\cite{Schaefer:2006sr,Herbst:2010rf,Skokov:2010wb}.

The most difficult problem within these model investigations is the
inclusion of the quantum back-reaction of the matter sector to the
gluonic sector. This problem has been attacked in Ref.
\cite{Schaefer:2007pw} which leads to a flavor and chemical potential
dependence of the (de-)confinement transition temperature. This first
perturbative estimate of Ref. \cite{Schaefer:2007pw} has been
confirmed by first-principle QCD calculations within the FRG approach
at real and imaginary chemical potentials \cite{Braun:2009gm} and also
by constraining PNJL model results with those in the statistical
model, see \cite{Fukushima:2010bq} for a recent review. This already
demonstrates that even with QCD-based effective models combined with
functional approaches valuable information about the QCD phase diagram
can be provided and certain scenarios can be excluded. As an example
see \cite{Nakano:2009ps} concerning the expected focusing of
isentropic trajectories in the vicinity of a critical endpoint in the
phase diagram. In this sense these models can be understood as
controlled approximations to full dynamical QCD.

Nevertheless, treating the QCD Green's functions directly is, of
course, very desireable. In the next two subsections two types of
investigations aiming at a direct calculation of the QCD Green's
function within functional methods will be described.

\subsection{$\mathbf {T\not = 0}$ and $\mathbf {\mu =0}$ case}

At $ {T\not = 0}$ the heat bath provides a prefered rest frame.
Accordingly, the Green's functions will not only depend on the Lorentz
invariant combinations of energy and momenta. This, by itself, makes
the solution of functional equations much more involved. In addition,
the number of functions which specify uniquely a given Green's
function increases. An exception is the ghost propagator which still
can be described by only one function. However, for the gluon
propagator the one renormalization function splits into two,
\begin{equation}
D_{\mu \nu}^{\rm{Gluon}}=\frac{{Z_T(p^2)}}{p^2}
P^T_{\mu \nu}(p) + \frac{{Z_L(p^2)}}{p^2} P^L_{\mu \nu}(p) ,
\end{equation}
according to transverse and longitudinal projections to the heat bath.
For further discussions it is important to note that the corresponding
contributions are essentially chromomagnetic and chromoelectric in
nature. For the quark propagator also the renormalization function
splits into two such that the quark propagator is described by three
functions\footnote{By symmetries a structure of the type
  ${\gamma_i} p_i \gamma_4 \omega_p {D(p)} $ would be allowed which,
  however, has turned out so far in all investigated cases as
  negligibly small.}
\begin{equation}
S(p)=\frac1{ -i {\gamma_i}  p_i {A(p)} -i
\gamma_4 \omega_p {C(p)} + {B(p)}} .
\label{ABCdef} 
\end{equation}

At $T=0$ one has of course $Z_T=Z_L=Z$ and $C=A$ which are functions of $p^2$
 only with $p=(\omega_p, \vec p)$. At $T\not =0$ we deal with six propagator
functions depending on  $\vec p^2$ and $\omega_p$. At $T\to \infty$ one obtains
a three-dimensional Yang-Mills theory plus a scalar field originating from
the $A_4$.

At any $T$ the chromomagnetic part of the gluon propagator keeps
positivity violating \cite{Maas:2004se,Cucchieri:2007ta,Maas:2005hs}.
This clearly indicates a kind of partial gluon confinement at any
temperature (a more detailed discussion may be found in
\cite{Lichtenegger:2008mh} and references therein). The results of
these investigations relate also to the non-perturbative origin of
chromomagnetic screening, whatever the temperature is. This is not at
all surprising because three-dimensional Yang-Mills theory is
confining (and thus the infrared behaviour of the gluon propagator is
genuinely non-perturbative), one has an area law for the spatial
Wilson loop, and the Coulomb string tension is non-vanishing at any
$T$ \cite{Greensite:2004ke}.

\begin{figure}
\centerline{
\resizebox{0.95\columnwidth}{!}{\includegraphics{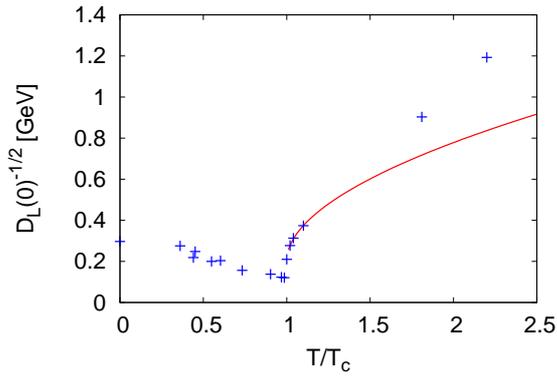}}
}
\caption{Displayed is the Landau gauge chromoelectric screening mass 
scale with a fit to the critical exponent near the critical point. The 
data points are taken from Ref.\ \cite{Fischer:2010fx}.}
\label{ScreenMass}       
\end{figure}

\begin{figure}
\centerline{
\resizebox{0.95\columnwidth}{!}{\includegraphics{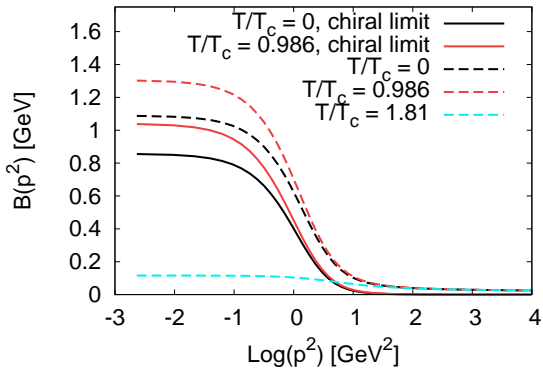}}
}
\caption{The Landau gauge scalar quark function $B(p^2)$ (see Eq.\
(\ref{ABCdef})) for 
different temperatures in the chiral limit (solid lines) 
and for a typical light-quark 
current quark mass (dashed lines).}
\label{ScalQuarkProp}       
\end{figure}

\begin{figure}
\centerline{
\resizebox{0.95\columnwidth}{!}{\includegraphics{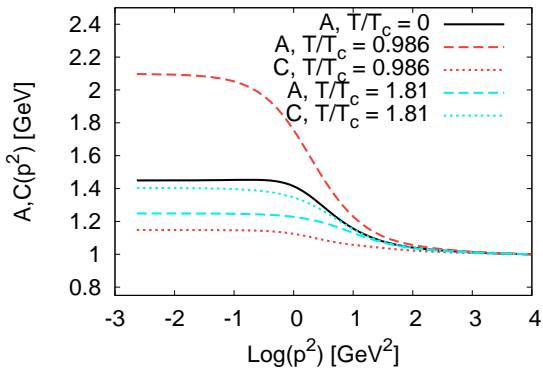}}
}
\caption{The Landau gauge quark renormalization functions $A(p^2)$ and
  $C(p^2)$ (see Eq.\ (\ref{ABCdef})) for different temperatures in the
  chiral limit.}
\label{QuarkProp}       
\end{figure}

To calculate the quark propagator we solve its DSE with the
finite-temperature gluon propagator and quark-gluon vertex as input.
In our corresponding first calculations we follow Ref.\
\cite{Fischer:2010fx} where a fit to the lattice gluon propagator and
an ansatz for the quark-gluon vertex has been used. As already stated
the chromomagnetic part of the gluon propagator is not at all
influenced by the phase transition. On the other hand, the
chromoelectric part may serve as an order parameter. In Fig.\
\ref{ScreenMass} one sees very clearly the drastice change in the
infrared value of the chromoelectric gluon propgator, $D_L(0)$, at
$T=T_c=$277 MeV. The inverse of the square root of $D_L(0)$ is the
chromoelectric screening mass scale, and from the fit in Fig.\
\ref{ScreenMass} we obtain a critical exponent of approximately 0.53.

The resulting quark propagator functions from the corresponding DSE
are displayed in Figs.\ \ref{ScalQuarkProp} and \ref{QuarkProp}. The
scalar function $B(p^2)$ is the one reflecting D$\chi$SB.\footnote{For
  conceptional clarity we prefer to discuss the function $B(p^2)$ and
  not one of the mass functions $M(p^2)=B(p^2)/A(p^2)$ or
  $M(p^2)=B(p^2)/C(p^2)$, respectively.} Correspondingly, in the
chiral limit it vanishes identically above $T_c$ thereby signaling the
chiral phase transition. Of course, this transition becomes a
crossover for non-vanishing current quark mass, and the function $B$
is slowly varying and mostly driven by the current quark mass. A not
anticipated behaviour is the fact that $B(0)$ raises with temperature
up to $T_c$. But Fig.\ \ref{ScreenMass} makes evident why this
happens: The infrared value of the chromoelectric part of the gluon
propagator increases with $T$ up to $T_c$ before it then sharply
decreases.

Putting as usual antiperiodic boundary conditions for quarks, the
chiral-limit quark condensate (being essentially the functional trace
over the quark propagator) serves as an order parameter. However,
playing with the boundary conditions allows to link confinement with
spectral properties of the Dirac operator \cite{Gattringer:2006ci}.
Correspondingly, in Ref.\ \cite{Fischer:2010fx} also the dual quark
condensate and the dressed Polykaov loop
\cite{Gattringer:2006ci,Synatschke:2007bz,Synatschke:2008yt} have been
calculated.\footnote{For corresponding $T=0$ lattice and functional
  equations see Refs.\ \cite{Bilgici:2008qy,Fischer:2009wc}. } The
trick consists in setting $U(1)$-valued boundary conditions for the
quark field in ``temporal'' direction:
\begin{equation}
q(\vec x, \beta =1/T) = e^{i{\varphi}} q(\vec x, 0),
\end{equation}
which amounts to an effective shift of Matsubara frequencies,
\begin{equation}
\omega_n = 2\pi T (n+ {\varphi}/2\pi ).
\end{equation}
The generalized condensate $\langle \bar q q \rangle _ {{\varphi}}$ 
at fixed $\varphi$ is the corresponding expectation value of the 
Dirac operator, and it is expandable in a 
geometric series containing loops of link variables with
increasing winding number. 
One can then define the dual (Gattringer) condensate as 
\begin{equation}
\Sigma_\nu = - \int _0 ^{2\pi} 
\frac {d{{\varphi}}}{2\pi} 
e^{-i\nu {{\varphi}}} \langle \bar q q \rangle _
{{\varphi}}
\end{equation}
where $\nu=1$ projects out winding-number-one loops, the so-called
dressed Polyakov loops \cite{Gattringer:2006ci}. It is an order
parameter for center symmetry breaking and confinement.

The aim of our future investigations is now to calculate the dual
quark condensate and the dressed Polyakov loop with an improved input.
Substituting the parameterized gluon propagator by recent results of
functional equations is hereby more a matter of convenience and
numerical precision than of fundamental progress. However, an (at
least partially) self-consistent determination of the quark propagator
and the quark-gluon vertex together, generalizing the work of
Ref.~\cite{Alkofer:2008tt} to non-vanishing temperatures, would be a
decisive step towards a first-principle calculation of
finite-temperature quark properties in continuum QCD.

At the end of this subsection we want to point out that one can define
a Polyakov-loop related confinement criterium from infrared exponents
\cite{Braun:2007bx}. Together with an FRG calculation of the Polyakov
potential this allows one to calculate the transition temperature. In
pure Yang-Mills theory one finds hereby for SU(2) a second-order phase
transition lying in the Ising universality class and a first-order
transition for all SU(N$_c\ge$3), Sp(2) and for the E(7) group
\cite{Braun:2010cy}.

\subsection{Color superconducting phase}

\begin{figure}
\centerline{\resizebox{0.95\columnwidth}{!}{\includegraphics{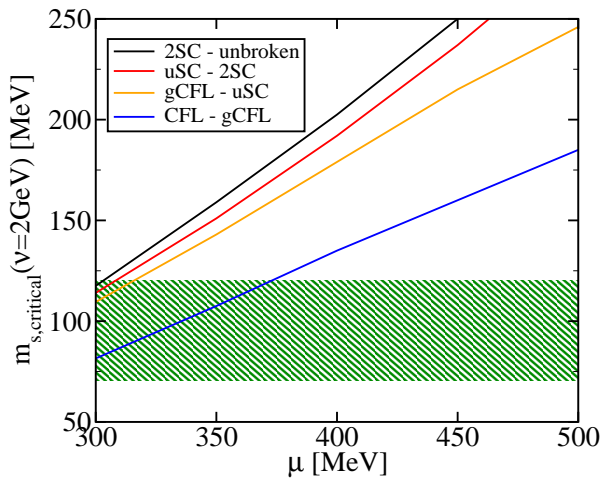}}}
\caption{Color superconducting phases in the $m_s-\mu$ plane.}
\label{CSCphases}       
\end{figure}

Of course, it is quite obvious to exploit the absence of the sign
problem in functional approaches to extend the studies to
non-vanishing density. Nevertheless, the introduction of a chemical
potential into functional equations leads to a significant
complication of the quark propagator parameterization, usually done in
the Nambu-Gorkov formalism then, see {\it e.g.\/} Refs.\
\cite{Nickel:2006vf,Nickel:2006kc,Nickel:2008ef} and references
therein. It has to be noted that as long as one uses an Abelian-type
model for the quark-gluon vertex these difficulties are merely
technical obstacles which can be overcome by combining computer
algebra with numerical techniques. Nevertheless, including medium
modifications by quarks (as done {\it e.g.\/} in Refs.\
\cite{Nickel:2006vf,Nickel:2006kc,Nickel:2008ef}) and employing the
results of functional equations or a combined phenomenological and
lattice fit to the gluon propagator (as {\it e.g.\/} in Ref.\
\cite{Bhagwat:2003vw}) one obtains a quite astonishing but robust
result: Restricting oneself to translationally invariant phases one
concludes that for a realistic renormalized strange quark mass for all
chemical potentials above the phase transition, quark matter is in the
colour-flavour locked colour-super\-conducting phase. In Fig.\
\ref{CSCphases} a resulting phase diagram in the plane of renormalized
strange quark mass and quark chemical potential is given.\footnote{To
  obtain this figure electric and colour neutrality have been imposed,
  but this is, however, not decisive for the main result.}
The shaded band in the figure indicates the value of the strange quark mass
as given by the Particle Data Group, the different lines are phase
separation lines from colour-flavour-locked to gapless
colour-flavour-locked (lowest line), from gapless
colour-flavour-locked to the uSC phase (second-lowest line), and so
on.

In addition, these calculations demonstrate that there are huge
deviations of the gap functions as compared to those extrapolated from
the weak-coupling result up to che\-mical potentials of the order of
several GeV. Even at such large chemical potentials perturbation
theory quantitatively fails. Furthermore, the light quarks screen
interaction also in the strange quark sector, an effect, which is not
seen in most models.

A calculation of the dressed Polyakov loop at finite chemical potential will
certainly shed more light on the nature of the finite-density phase transition,
and therefore a corresponding generalization of the work described in Ref.\ 
\cite{Nickel:2006vf,Nickel:2006kc,Nickel:2008ef} is in progress \cite{Davor}. 
However, it has to be stated clearly here that the main aspects of the QCD phase
transition at finite density will not be uncovered by functional methods if we
will not get more information on the coupling of quarks to gluons also in this
region of the phase diagram. In the case of a DSE study this especially 
includes more knowledge about the  quark-gluon vertex

\section{Summary and Outlook: What may we expect?}

The last years have seen a quite dramatic increase in our knowledge on the
Landau gauge QCD Green's functions at vanishing temperatures and densities. It
has become evident that the propagation of transverse gluons is infrared
suppressed, and that positivity is violated for transverse gluons. It has turned
out that ghosts are the long-range ``degrees of freedom'', and concentrating on
the main point one may say that in Landau gauge QCD, ghosts and the restriction
to the Gribov horizon ({\it i.e.\/} the
correlations introduced by gauge-fixing uniquely) are related to the origin of
confinement \cite{Zwanziger:2003cf,Alkofer:2000wg}.

It cannot be emphasized enough that dynamical chiral symmetry breaking
does not only lead to the generation of constituent quark masses but
also to scalar-type couplings between quarks and gluons. There are
indications that the related scalar confinement potential is even
larger than the vector component \cite{Alkofer:2008tt}. 
We also remark that the axial
anomaly is encrypted in the infrared behaviour of the Green's
functions in the scaling solution \cite{Alkofer:2008et}.

Having set the stage, one is ready to continue to either non-vanishing
temperatures or/and densities. Firstly, one sh\-ould note that the
chromomagnetic part of the gluon propagator does not see the phase
transition at all. In particular, one has positivity violation (and
thus partial gluon confinement) at any temperature. Secondly, the
infrared region of the chromoelectric part of the gluon propagator
displays very nicely the phase transition \cite{Fischer:2010fx}. 
Needless to say, that in
the chiral limit the quark propagator changes also accordingly due to
the chiral symmetry restoration at the critical temperature.

The results of the quark propagator DSE at non-vani\-shing densities provide
evidence that for a realistic renormalized strange quark mass for all values of
the chemical potentials above the phase transition, quark matter is in the
colour-flavour locked colour-superconducting phase  \cite{Nickel:2008ef}.
However, here we have to keep in mind that these investigations are restricted
to translationally invariant phases.

As we know from recent studies within functional approaches at $T=0$
and $\mu=0$ the dressing of the quark-gluon vertex is of utter
importance. Therefore we will concentrate in future on investigations
of this function at non-vani\-shing temperatures and densities.
Hereby, the dual qua\-rk condensate and the dressed Polyakov loop 
\cite{Bilgici:2008qy} will
be important quanitities in the studies of the QCD phase diagram.
Although it may sound very ambitious today we want to emphasize that
the goal is to calculate thermodynamic observables of QCD at all
physically relevant temperatures and chemical potentials.

\section*{Acknowledgments}
RA thanks the organizers of this highly interesting workshop, in particular
Tam\'as Bir\'o, for the invitation. 

We are grateful to Christian Fischer, Davor Horvatic, Axel Maas,  Dominik
Nickel, Jan Martin Pawlowski, Lorenz von Smekal, 
and Jochen Wambach for helpful discussions. 

We thank Jan Martin Pawlowski for a critical reading of the manuscript.

This work was supported by  the Austrian Science Fund FWF
within the Doctoral Program No.\ W1203,
and in part by the European Union (HadronPhysics2 project
``Study of strongly-interacting matter'').


\begin{thebibliography}{99}
\bibitem{Nicmorus:2010sd}
  D.~Nicmorus, G.~Eichmann, R.~Alkofer,
  %``Delta and Omega electromagnetic form factors in a Dyson-Schwinger/Bethe-Salpeter approach,''
  Phys. Rev.  {\bf D82} (2010)  114017
  [arXiv:1008.3184 [hep-ph]].
  %%CITATION = ARXIV:1008.3184;%%
  
\bibitem{SanchisAlepuz:2010in}
  H.~Sanchis-Alepuz {\it et al.},
  %R.~Alkofer, G.~Eichmann and S.~Villalba-Chavez,
  %``On baryon properties from a covariant Faddeev approach,''
  PoS {\bf LC2010} (2010) 018
  [arXiv: 1010.6183 [hep-ph]].
  %%CITATION = POSCI,LC2010,018;%%

%
\bibitem{Pawlowski:2010ht}
  J.~M.~Pawlowski,
  %``The QCD phase diagram: Results and challenges,''
    arXiv:1012.5075 [hep-ph].
  %%CITATION = ARXIV:1012.5075;%%


%
\bibitem{Alkofer:2008bs}
  R.~Alkofer {\it et al.},
  %C.~S.~Fischer, M.~Q.~Huber, F.~J.~Llanes-Estrada and K.~Schwenzer,
  %``Confinement and Green functions in Landau-gauge QCD,''
  PoS  {\bf CONFINEMENT8} (2008) 019
  [arXiv:0812.2896 [hep-ph]].
  %%CITATION = POSCI,CONFINEMENT8,019;%%

%
\bibitem{Sternbeck:2006cg}
  A.~Sternbeck  {\it et al.},
  %E.~M.~Ilgenfritz, M.~Muller-Preussker, A.~Schiller and I.~L.~Bogolubsky,
  %``Lattice study of the infrared behavior of QCD Green's functions in Landau
  %gauge,''
  PoS {\bf LAT2006} (2006) 076
  [arXiv: hep-lat/0610053].
  %%CITATION = POSCI,LAT2006,076;%%

%
\bibitem{von Smekal:1997is}
  L.~von Smekal, R.~Alkofer and A.~Hauck,
  %``The infrared behavior of gluon and ghost propagators in Landau gauge
  %QCD,''
  Phys.\ Rev.\ Lett.\  {\bf 79} (1997) 3591
  [arXiv:hep-ph/9705242].
  %%CITATION = PRLTA,79,3591;%%

%
\bibitem{Fischer:2002hna}
  C.~S.~Fischer and R.~Alkofer,
  %``Infrared exponents and running coupling of SU(N) Yang-Mills theories,''
  Phys.\ Lett.\  B {\bf 536} (2002) 177
  [arXiv:hep-ph/0202202].
  %%CITATION = PHLTA,B536,177;%%

%
\bibitem{Pawlowski:2003hq}
  J.~M.~Pawlowski {\it et al.},
  %D.~F.~Litim, S.~Nedelko and L.~von Smekal,
  %``Infrared behaviour and fixed points in Landau gauge QCD,''
  Phys.\ Rev.\ Lett.\  {\bf 93} (2004) 152002
  [arXiv: hep-th/0312324].
  %%CITATION = PRLTA,93,152002;%%

\bibitem{Fischer:2008uz}
  C.~S.~Fischer, A.~Maas and J.~M.~Pawlowski,
  %``On the infrared behavior of Landau gauge Yang-Mills theory,''
  Annals Phys.\  {\bf 324} (2009) 2408
  [arXiv:0810.1987 [hep-ph]].
  %%CITATION = APNYA,324,2408;%%
 

\bibitem{Taylor:1971ff}
J.~C. Taylor, {Nucl. Phys.} {\bf B33} (1971) 436.
%%CITATION = NUPHA,B33,436;%%.

\bibitem{Lerche:2002ep}
C.~Lerche and L.~von Smekal,
%``On the infrared exponent for gluon and ghost propagation in Landau  gauge QCD,''
Phys.\ Rev.\ D {\bf 65}  (2002) 125006
 [arXiv:hep-ph/0202194].
 %%CITATION = HEP-PH 0202194;%%

\bibitem{Cucchieri:2004sq}
A.~Cucchieri, T.~Mendes, and A.~Mihara {JHEP} {\bf 12} (2004) 012
[arXiv:hep-lat/0408034].
%%CITATION = HEP-LAT 0408034;%%.

\bibitem{Schleifenbaum:2004id}
W.~Schleifenbaum {\it et al.},
%A.~Maas, J.~Wambach, R.~Alkofer,
{Phys. Rev. }{\bf  D72} (2005) \newline 014017
[arXiv:hep-ph/0411052].
%%CITATION = PHRVA,D72,014017;%%

\bibitem{Watson:2001yv}
  P.~Watson and R.~Alkofer,
  %``Verifying the Kugo-Ojima confinement criterion in Landau gauge QCD,''
  Phys.\ Rev.\ Lett.\  {\bf 86} (2001) 5239
  [arXiv:hep-ph/0102332].
  %%CITATION = PRLTA,86,5239;%%

%
\bibitem{Alkofer:2000wg}
  R.~Alkofer and L.~von Smekal,
  %``The infrared behavior of QCD Green's functions: Confinement, dynamical
  %symmetry breaking, and hadrons as relativistic bound states,''
  Phys.\ Rept.\  {\bf 353} (2001) 281
  [arXiv:hep-ph/0007355].
  %%CITATION = PRPLC,353,281;%%

%\cite{Braun:2009gm}
\bibitem{Braun:2009gm}
  J.~Braun {\it et al.}, 
  % L.~M.~Haas, F.~Marhauser and J.~M.~Pawlowski,
  Phys.\ Rev.\ Lett. (2011) {\it in press} 
  %``On the relation of quark confinement and chiral symmetry breaking,''
  [arXiv:0908.0008 [hep-ph]].
  %%CITATION = ARXIV:0908.0008;%%


\bibitem{Alkofer:2004it}
  R.~Alkofer, C.~S.~Fischer and F.~J.~Llanes-Estrada,
  % ``Vertex functions and infrared fixed point in Landau gauge SU(N)  Yang-Mills
  %theory,''
  Phys.\ Lett.\ B {\bf 611} (2005)  279
  [arXiv:hep-th/0412330].
  %%CITATION = PHLTA,B611,279;%%

\bibitem{Huber:2007kc}
  M.~Q.~Huber {\it et al.},
  %R.~Alkofer, C.~S.~Fischer and K.~Schwenzer,
  %``The infrared behavior of Landau gauge Yang-Mills theory in d=2, 3 and 4
  %dimensions,''
  Phys.\ Lett.\  B {\bf 659} (2008) 434
 [arXiv:0705.3809 [hep-ph]].
  %%CITATION = PHLTA,B659,434;%%


%
\bibitem{Alkofer:2008nt}
  R.~Alkofer, M.~Q.~Huber and K.~Schwenzer,
  %``Algorithmic derivation of Dyson-Schwinger Equations,''
  Comput.\ Phys.\ Commun.\  {\bf 180} (2009) 965
  [arXiv:0808.2939 [hep-th]].
  %%CITATION = CPHCB,180,965;%%

%
\bibitem{Fischer:2006vf}
  C.~S.~Fischer and J.~M.~Pawlowski,
  %``Uniqueness of infrared asymptotics in Landau gauge Yang-Mills theory,''
  Phys.\ Rev.\  D {\bf 75} (2007) 025012
  [arXiv:hep-th/0609009].
  %%CITATION = PHRVA,D75,025012;%%

%
\bibitem{Fischer:2009tn}
  C.~S.~Fischer and J.~M.~Pawlowski,
  %``Uniqueness of infrared asymptotics in Landau gauge Yang-Mills theory II,''
  Phys.\ Rev.\  D {\bf 80} (2009) 025023
  [arXiv:0903.2193 [hep-th]].
  %%CITATION = PHRVA,D80,025023;%%

\bibitem{Szczepaniak:2001rg}
  A.~P.~Szczepaniak and E.~S.~Swanson,
  %``Coulomb gauge QCD, confinement, and the constituent representation,''
  Phys.\ Rev.\  D {\bf 65} (2002) 025012
  [arXiv:hep-ph/0107078].
  %%CITATION = PHRVA,D65,025012;%%


\bibitem{Aguilar:2008xm}
  A.~C.~Aguilar, D.~Binosi and J.~Papavassiliou,
  %``Gluon and ghost propagators in the Landau gauge: Deriving lattice results
  %from Schwinger-Dyson equations,''
  Phys.\ Rev.\  D {\bf 78} (2008) 025010
  [arXiv:0802.1870 [hep-ph]].
  %%CITATION = PHRVA,D78,025010;%%

\bibitem{Boucaud:2008ky}
 P.~Boucaud {\it et al.},
%  P.~Boucaud, J.~P.~Leroy, A.~Le Yaouanc, J.~Micheli, O.~Pene and J.~Rodriguez-Quintero,
  %``On the IR behaviour of the Landau-gauge ghost propagator,''
  JHEP {\bf 0806} (2008) 099
  [arXiv: 0803.2161 [hep-ph]].
  %%CITATION = JHEPA,0806,099;%%

\bibitem{Alkofer:2008jy}
  R.~Alkofer, M.~Q.~Huber and K.~Schwenzer,
  %``Infrared singularities in Landau gauge Yang-Mills theory,''
  Phys.\ Rev.\  D {\bf 81} (2010) 105010
  [arXiv:0801.2762 [hep-th]].
  %%CITATION = PHRVA,D81,105010;%%

\bibitem{Huber:2009wh}
  M.~Q.~Huber, K.~Schwenzer and R.~Alkofer,
  %``On the infrared scaling solution of SU(N) Yang-Mills theories in the
  %maximally Abelian gauge,''
  Eur.\ Phys.\ J.\  {\bf C68 } (2010)  581-600
  [arXiv:0904.1873 [hep-th]].
  %%CITATION = ARXIV:0904.1873;%%


\bibitem{Sternbeck:2008mv}
  A.~Sternbeck and L.~von Smekal,
  %``Infrared exponents and the strong-coupling limit in lattice Landau gauge,''
  Eur.\ Phys.\ J.\  C {\bf 68} (2010) 487
  [arXiv:0811.4300 [hep-lat]].
  %%CITATION = EPHJA,C68,487;%%

\bibitem{Maas:2009ph}
  A.~Maas {\it et al.},
  %J.~M.~Pawlowski, D.~Spielmann, A.~Sternbeck and L.~von Smekal,
  %``Strong-coupling study of the Gribov ambiguity in lattice Landau gauge,''
  Eur.\ Phys.\ J.\  C {\bf 68} (2010) 183
  [arXiv: 0912.4203 [hep-lat]].
  %%CITATION = EPHJA,C68,183;%%

\bibitem{Cucchieri:2009zt}
  A.~Cucchieri and T.~Mendes,
  %``Landau-gauge propagators in Yang-Mills theories at beta = 0: massive
  %solution versus conformal scaling,''
  Phys.\ Rev.\  D {\bf 81} (2010) 016005
  [arXiv:0904.4033 [hep-lat]].
  %%CITATION = PHRVA,D81,016005;%%

\bibitem{Maas:2009se}
  A.~Maas,
  %``Constructing non-perturbative gauges using correlation functions,''
  Phys.\ Lett.\  B {\bf 689} (2010) 107
  [arXiv:0907. 5185 [hep-lat]];
  %%CITATION = PHLTA,B689,107;%%

%
\bibitem{Zwanziger:2003cf}
  D.~Zwanziger,
  %``Non-perturbative Faddeev-Popov formula and infrared limit of QCD,''
  Phys.\ Rev.\  D {\bf 69} (2004) 016002
  [arXiv: hep-ph/0303028].
  %%CITATION = PHRVA,D69,016002;%%



\bibitem{Alkofer:2003jj}
  R.~Alkofer {\it et al.},
  %W.~Detmold, C.~S.~Fischer, P.~Maris,
  %``Analytic properties of the Landau gauge gluon and quark propagators,''
  Phys.\ Rev.\  D {\bf 70} (2004) 014014;
  [arXiv:hep-ph/0309077];
 %%CITATION = PHRVA,D70,014014;%%
%\cite{Alkofer:2003jk}
%\bibitem{Alkofer:2003jk}
 %  R.~Alkofer, W.~Detmold, C.~S.~Fischer and P.~Maris,
  %``Analytic structure of the gluon and quark propagators in Landau gauge
  %QCD,''
  Nucl.\ Phys.\ Proc.\ Suppl.\  {\bf 141} (2005) 122
  [arXiv:hep-ph/0309078].
  %%CITATION = NUPHZ,141,122;%%

\bibitem{Bowman:2007du}
  P.~O.~Bowman {\it et al.},
  %``Scaling behavior and positivity violation of the gluon propagator in full
  %QCD,''
  Phys.\ Rev.\ D {\bf 76} (2007) 094505
  [arXiv:hep-lat/0703022].
  %%CITATION = HEP-LAT/0703022;%%

\bibitem{Alkofer:2006gz}
  R.~Alkofer, C.~S.~Fischer and F.~J.~Llanes-Estrada,
  %``Dynamically induced scalar quark confinement,''
  Mod.\ Phys.\ Lett.\   {\bf A23} (2008) 1105
  [arXiv:hep-ph/ 0607293].
  %%CITATION = MPLAE,A23,1105;%%


\bibitem{Alkofer:2008tt}
  R.~Alkofer {\it et al.},
  %C.~S.~Fischer, F.~J.~Llanes-Estrada, and K.~Schwenzer,
  %``The quark-gluon vertex in Landau gauge QCD: Its role in dynamical chiral
  %symmetry breaking and quark confinement,''
  Annals Phys.\  {\bf 324} (2009) 106
  [arXiv: 0804.3042 [hep-ph]].
  %%CITATION = APNYA,324,106;%%


%
\bibitem{Alkofer:2008et}
  R.~Alkofer, C.~S.~Fischer and R.~Williams,
  %``U_A(1) anomaly and eta' mass from an infrared singular quark-gluon
  %vertex,''
  Eur.\ Phys.\ J.\  A {\bf 38} (2008) 53
  [arXiv:0804.3478 [hep-ph]].
  %%CITATION = EPHJA,A38,53;%%

\bibitem{Kogut:1973ab}
  J.~B.~Kogut and L.~Susskind,
  %``Quark Confinement And The Puzzle Of The Ninth Axial Current,''
  Phys.\ Rev.\  D {\bf 10} (1974) 3468.
  %%CITATION = PHRVA,D10,3468;%%


%\cite{Ratti:2005jh}
\bibitem{Ratti:2005jh}
  C.~Ratti, M.~A.~Thaler and W.~Weise,
  %``Phases of QCD: Lattice thermodynamics and a field theoretical model,''
  Phys.\ Rev.\  D {\bf 73} (2006) 014019
  [arXiv:hep-ph/0506234].
  %%CITATION = PHRVA,D73,014019;%%


%\cite{Schaefer:2007pw}
\bibitem{Schaefer:2007pw}
  B.-J.~Schaefer, J.M.~Pawlowski and J.~Wambach,
  %``The Phase Structure of the Polyakov--Quark-Meson Model,''
  Phys.\ Rev.\  {\bf D76} (2007) 074023
  [arXiv:0704.3234 [hep-ph]].
  %%CITATION = PHRVA,D76,074023;%%

%\cite{Schaefer:2009ui}
\bibitem{Schaefer:2009ui}
  B.-J.~Schaefer, M.~Wagner and J.~Wambach,
  %``Thermodynamics of (2+1)-flavor QCD: Confronting Models with Lattice
  %Studies,''
  Phys.\ Rev.\  D {\bf 81}, 074013 (2010)
  [arXiv:0910.5628 [hep-ph]].
  %%CITATION = PHRVA,D81,074013;%%
%\cite{Schaefer:2008hk}
%\bibitem{Schaefer:2008hk}
  B.-J.~Schaefer and M.~Wagner,
  %``The three-flavor chiral phase structure in hot and dense QCD matter,''
  Phys.\ Rev.\  D {\bf 79}, 014018 (2009)
  [arXiv:0808.1491 [hep-ph]].
  %%CITATION = PHRVA,D79,014018;%%


%\cite{Schaefer:2006sr}
\bibitem{Schaefer:2006sr}
  B.-J.~Schaefer and J.~Wambach,
  %``Renormalization group approach towards the QCD phase diagram,''
  Phys.\ Part.\ Nucl.\  {\bf 39} (2008) 1025
  [arXiv:hep-ph/0611191].
  %%CITATION = PPNUE,39,1025;%%


%\cite{Herbst:2010rf}
\bibitem{Herbst:2010rf}
  T.~K.~Herbst, J.~M.~Pawlowski and B.-J.~Schaefer,
  %``The phase structure of the Polyakov--quark-meson model beyond mean field,''
  Phys. Lett. B {\bf 696} (2011) 58 %-67
  [arXiv:1008.0081 [hep-ph]].
  %%CITATION = ARXIV:1008.0081;%%


%\cite{Skokov:2010wb}
\bibitem{Skokov:2010wb}
 V.~Skokov  {\it et al.},
%  V.~Skokov, B.~Stokic, B.~Friman and K.~Redlich,
  %``Meson fluctuations and thermodynamics of the Polyakov loop extended
  %quark-meson model,''
  Phys.\ Rev.\  C {\bf 82}, 015206 (2010)
  [arXiv:1004.2665 [hep-ph]].
  %%CITATION = PHRVA,C82,015206;%%



%\cite{Fukushima:2010bq}
\bibitem{Fukushima:2010bq}
  K.~Fukushima and T.~Hatsuda,
  %``The phase diagram of dense QCD,''
  Rept.\ Prog.\ Phys.\  {\bf 74} (2011) 014001
  [arXiv:1005.4814 [hep-ph]].
  %%CITATION = RPPHA,74,014001;%%

%\cite{Nakano:2009ps}
\bibitem{Nakano:2009ps}
  E.~Nakano, B.-J.~Schaefer, B.~Stokic, B.~Friman and K.~Redlich,
  %``Fluctuations and isentropes near the chiral critical endpoint,''
  Phys.\ Lett.\  B {\bf 682} (2010) 401
  [arXiv:0907.1344 [hep-ph]].
  %%CITATION = PHLTA,B682,401;%%

%\cite{Maas:2004se}
\bibitem{Maas:2004se}
  A.~Maas {\it et al.},
  %J.~Wambach, B.~Gruter and R.~Alkofer,
  %``High-temperature limit of Landau-gauge Yang-Mills theory,''
  Eur.\ Phys.\ J.\  C {\bf 37} (2004) 335
  [arXiv: hep-ph/0408074].
  %%CITATION = EPHJA,C37,335;%%

%\cite{Cucchieri:2007ta}
\bibitem{Cucchieri:2007ta}
  A.~Cucchieri, A.~Maas and T.~Mendes,
  %``Infrared properties of propagators in Landau-gauge pure Yang-Mills   theory
  %at finite temperature,''
  Phys.\ Rev.\  D {\bf 75} (2007) 076003
  [arXiv:hep-lat/0702022].
  %%CITATION = PHRVA,D75,076003;%%



%\cite{Maas:2005hs}
\bibitem{Maas:2005hs}
  A.~Maas, J.~Wambach and R.~Alkofer,
  %``The high-temperature phase of Landau-gauge Yang-Mills theory,''
  Eur.\ Phys.\ J.\  C {\bf 42} (2005) 93
  [arXiv:hep-ph/0504019].
  %%CITATION = EPHJA,C42,93;%%


%\cite{Lichtenegger:2008mh}
\bibitem{Lichtenegger:2008mh}
  K.~Lichtenegger and D.~Zwanziger,
  %``Nonperturbative contributions to the QCD pressure,''
  Phys.\ Rev.\  D {\bf 78} (2008) 034038
  [arXiv:0805.3804 [hep-ph]].
  %%CITATION = PHRVA,D78,034038;%%

%
\bibitem{Greensite:2004ke}
  J.~Greensite, S.~Olejnik and D.~Zwanziger,
  %``Coulomb energy, remnant symmetry, and the phases of non-Abelian gauge
  %theories,''
  Phys.\ Rev.\  D {\bf 69} (2004) 074506
  [arXiv:hep-lat/0401003].
  %%CITATION = PHRVA,D69,074506;%%


%\cite{Fischer:2010fx}
\bibitem{Fischer:2010fx}
  C.~S.~Fischer, A.~Maas and J.~A.~Muller,
  %``Chiral and deconfinement transition from correlation functions: SU(2) vs.
  %SU(3),''
  Eur.\ Phys.\ J.\  C {\bf 68} (2010) 165
  [arXiv:1003.1960 [hep-ph]].
  %%CITATION = EPHJA,C68,165;%%

%\cite{Gattringer:2006ci}
\bibitem{Gattringer:2006ci}
  C.~Gattringer,
  %``Linking confinement to spectral properties of the Dirac operator,''
  Phys.\ Rev.\ Lett.\  {\bf 97} (2006) 032003
  [arXiv:hep-lat/0605018].
  %%CITATION = PRLTA,97,032003;%%

%\cite{Synatschke:2007bz}
\bibitem{Synatschke:2007bz}
  F.~Synatschke, A.~Wipf and C.~Wozar,
  %``Spectral sums of the Dirac-Wilson operator and their relation to the
  %Polyakov loop,''
  Phys.\ Rev.\  D {\bf 75} (2007) 114003
  [arXiv:hep-lat/0703018].
  %%CITATION = PHRVA,D75,114003;%%

%\cite{Synatschke:2008yt}
\bibitem{Synatschke:2008yt}
  F.~Synatschke, A.~Wipf and K.~Langfeld,
  %``Relation between chiral symmetry breaking and confinement in YM-theories,''
  Phys.\ Rev.\  D {\bf 77} (2008) 114018
  [arXiv:0803.0271 [hep-lat]].
  %%CITATION = PHRVA,D77,114018;%%

%\cite{Bilgici:2008qy}
\bibitem{Bilgici:2008qy}
  E.~Bilgici  {\it et al.},
  %F.~Bruckmann, C.~Gattringer and C.~Hagen,
  %``Dual quark condensate and dressed Polyakov loops,''
  Phys.\ Rev.\  D {\bf 77} (2008) 094007
  [arXiv:0801.4051 [hep-lat]].
  %%CITATION = PHRVA,D77,094007;%%

%\cite{Fischer:2009wc}
\bibitem{Fischer:2009wc}
  C.~S.~Fischer,
  %``Deconfinement phase transition and the quark condensate,''
  Phys.\ Rev.\ Lett.\  {\bf 103} (2009) 052003
  [arXiv:0904.2700 [hep-ph]].
  %%CITATION = PRLTA,103,052003;%%

%\cite{Braun:2007bx}
\bibitem{Braun:2007bx}
  J.~Braun, H.~Gies and J.~M.~Pawlowski,
  %``Quark Confinement from Color Confinement,''
  Phys.\ Lett.\  B {\bf 684} (2010) 262
  [arXiv:0708.2413 [hep-th]].
  %%CITATION = PHLTA,B684,262;%%

%\cite{Braun:2010cy}
\bibitem{Braun:2010cy}
  J.~Braun  {\it et al.},
  %A.~Eichhorn, H.~Gies and J.~M.~Pawlowski,
  %``On the Nature of the Phase Transition in SU(N), Sp(2) and E(7) Yang-Mills
  %theory,''
  Eur.\ Phys.\ J.\  C {\bf 70} (2010) 689
  [arXiv: 1007.2619 [hep-ph]].
  %%CITATION = EPHJA,C70,689;%%


%\cite{Nickel:2006vf}
\bibitem{Nickel:2006vf}
  D.~Nickel, J.~Wambach and R.~Alkofer,
  %``Color-superconductivity in the strong-coupling regime of Landau gauge
  %QCD,''
  Phys.\ Rev.\  D {\bf 73} (2006) 114028
  [arXiv:hep-ph/0603163].
  %%CITATION = PHRVA,D73,114028;%%


%\cite{Nickel:2006kc}
\bibitem{Nickel:2006kc}
  D.~Nickel, R.~Alkofer and J.~Wambach,
  %``On the unlocking of color and flavor in color-superconducting quark
  %matter,''
  Phys.\ Rev.\  D {\bf 74} (2006) 114015
  [arXiv:hep-ph/0609198].
  %%CITATION = PHRVA,D74,114015;%%


%\cite{Nickel:2008ef}
\bibitem{Nickel:2008ef}
  D.~Nickel, R.~Alkofer and J.~Wambach,
  %``Neutrality of the color-flavor-locked phase in a Dyson-Schwinger
  %approach,''
  Phys.\ Rev.\  D {\bf 77} (2008) 114010
  [arXiv:0802.3187 [hep-ph]].
  %%CITATION = PHRVA,D77,114010;%%

%\cite{Bhagwat:2003vw}
\bibitem{Bhagwat:2003vw}
  M.~S.~Bhagwat {\it et al.},
  %M.~A.~Pichowsky, C.~D.~Roberts and P.~C.~Tandy,
  %``Analysis of a quenched lattice-QCD dressed-quark propagator,''
  Phys.\ Rev.\  C {\bf 68} (2003) 015203
  [arXiv:nucl-th/0304003].
  %%CITATION = PHRVA,C68,015203;%%

\bibitem{Davor}
R.~Alkofer, D.\ Horvatic, B.-J.\ Schaefer, in preparation.


\end{thebibliography}
\end{document}